\documentclass[final,3p,twocolumn]{elsarticle}
\usepackage{graphicx}
\usepackage{amssymb}

\journal{Solid State Communications}
\begin{document}
\begin{frontmatter}



\title{Heavy-fermion superconductivity in Ce$_2$PdIn$_{8}$}


\author{D. Kaczorowski, D. Gnida, A.P. Pikul, and V.H. Tran}

\address{Institute of Low Temperature and Structure Research, Polish Academy of Sciences,
P. O. Box 1410, 50-950 Wroc{\l}aw 2, Poland}

\begin{abstract}
The compound Ce$_2$PdIn$_{8}$ is a recently discovered novel member of the series
Ce$_nT$In$_{3n+2}$, where $T$ = $d$-electron transition metal, and $n$ = 1 or 2. So far, only the
phases with $T$ = Co, Rh and Ir have been intensively studied for their unconventional
superconducting behaviors at low temperatures. By means of magnetic susceptibility, electrical
resistivity and heat capacity measurements we provide evidence that also Ce$_2$PdIn$_{8}$ has a
superconducting ground state with strong heavy-fermion character. The clean-limit
superconductivity sets in at $T_{\rm c}$ = 0.7 K at ambient pressure, likely at a verge of a
quantum phase transition that manifests itself in a form of distinct non-Fermi liquid features in
the bulk normal state characteristics.
\end{abstract}


\end{frontmatter}


\section{Introduction}

\begin{figure}[!ht]
\includegraphics[width=\columnwidth]{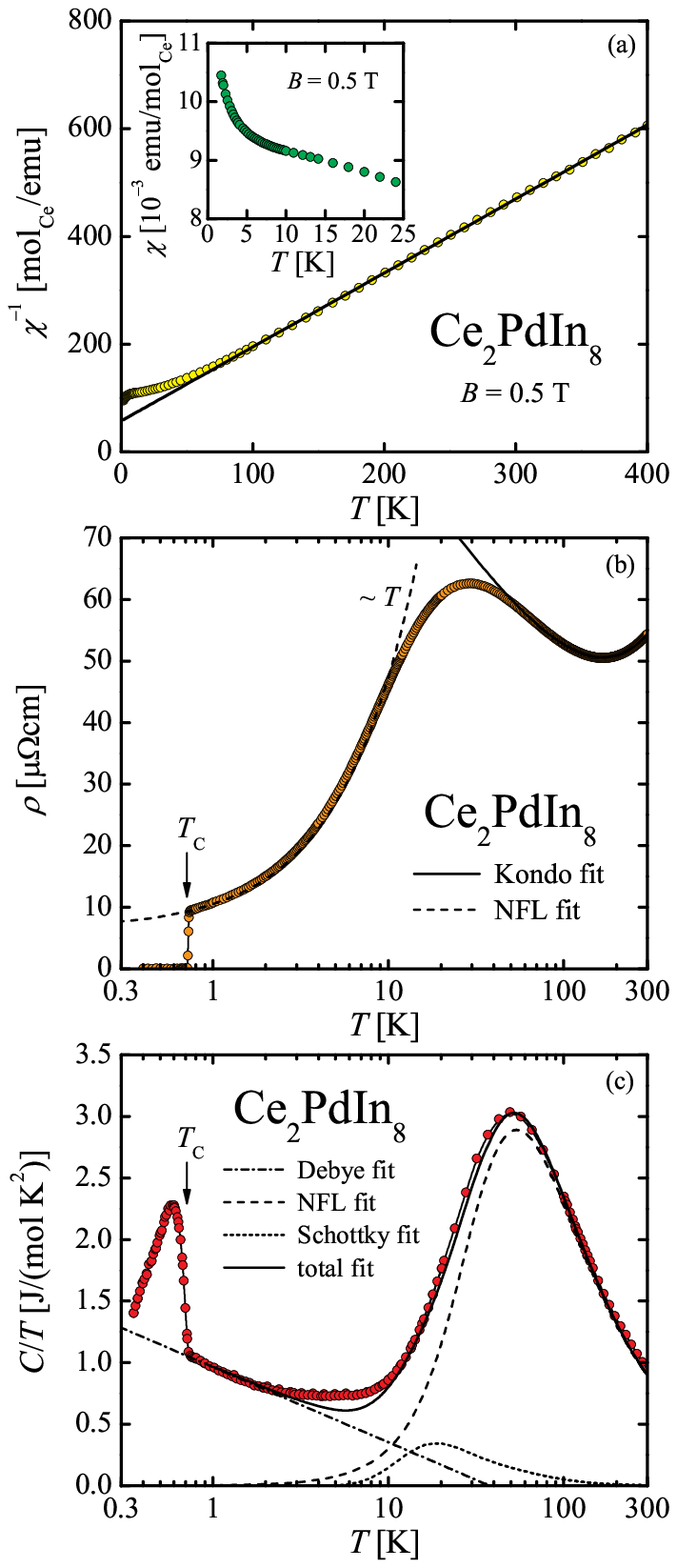}
\caption{(a) Temperature dependence of the inverse magnetic susceptibility of Ce$_2$PdIn$_8$
measured in a magnetic field of 0.5 T. The solid line represents the Curie-Weiss fit. The inset shows
the magnetic susceptibility at low temperatures. (b) Temperature variation of the electrical
resistivity of Ce$_2$PdIn$_8$. The lines stand for the fits discussed in the text. (c) Temperature
dependence of the specific heat over temperature ratio of Ce$_2$PdIn$_8$. The lines represent the
specific heat contributions discussed in the text.}
\end{figure}

\begin{figure}[!ht]
\includegraphics[width=\columnwidth]{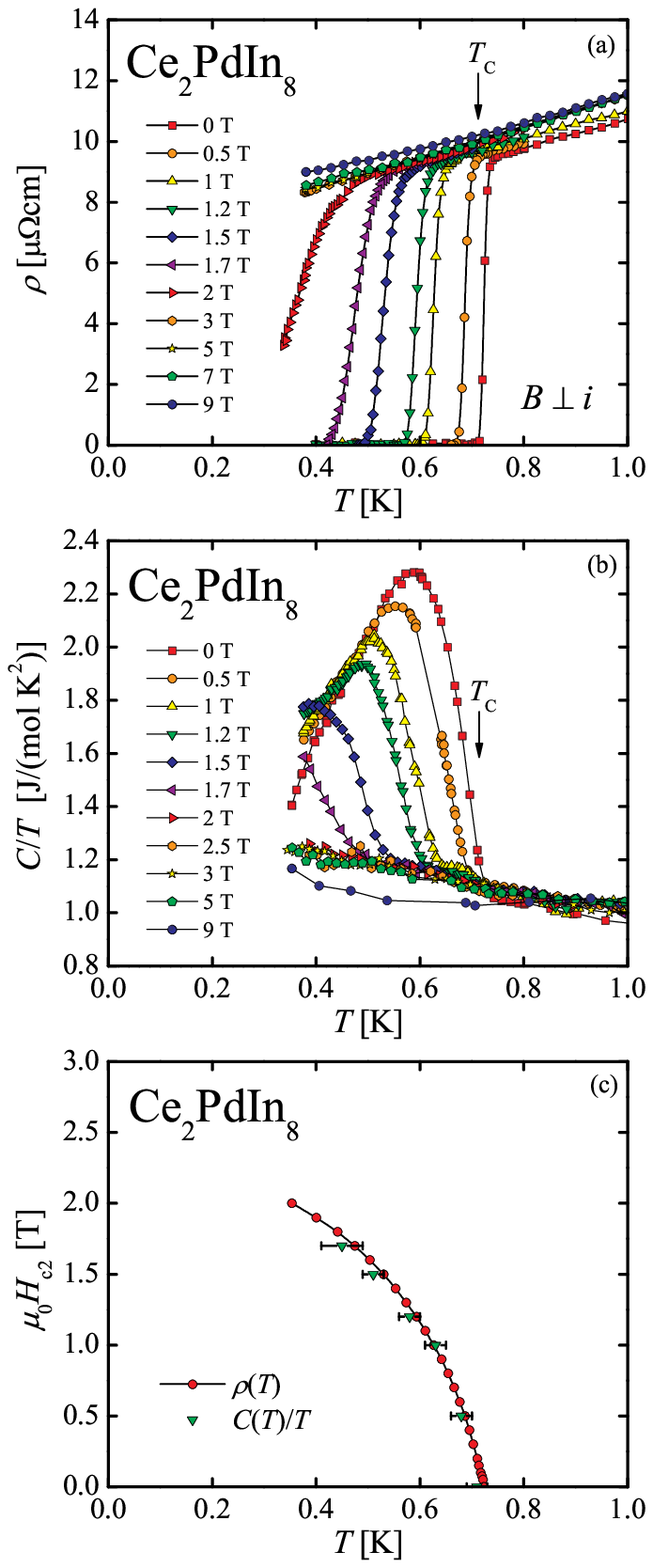}
\caption{Temperature dependencies of (a) the electrical resistivity (taken with the current $j$
= 0.1 mA) and (b) the specific heat over temperature ratio of Ce$_2$PdIn$_8$, measured in various
magnetic fields. (c) Temperature variation of the upper critical field in Ce$_2$PdIn$_8$,
determined from the data presented in panels (a) and (b).}
\end{figure}

Since the discovery of pressure-induced superconductivity in the heavy-fermion
antiferromagnet CeRhIn$_5$ \cite{heg_cerhin5}, the series of compounds Ce$T$In$_5$ ($T$ = Co,
Rh and Ir) has became one of the most intensively studied strongly correlated electron system. The
indides CeCoIn$_5$ and CeIrIn$_5$ have been characterized as ambient-pressure heavy-fermion
superconductors ($T_{\rm c}$ = 2.3 and 0.4 K, respectively) with paramagnetic normal state
properties \cite{pet_cecoin5,pet_ceirin5}. Moreover, in CeCoIn$_5$ a
Fulde-Farell-Larkin-Ovchinnikov phase has been observed in strong magnetic fields, in which the
electrons form Cooper pairs with nonzero total momentum \cite{FFLO}. In turn, in CeRhIn$_5$ the
heavy-fermion superconductivity emerges out of an antiferromagnetic state ($T_{\rm N}$ = 3.8 K at
ambient pressure) upon application of hydrostatic pressure ($T_{\rm c}$ = 2 K at $p$ = 1.6 GPa)
\cite{heg_cerhin5}. Similar properties have been found for the structurally closely related
phases Ce$_2T$In$_8$ with $T$ = Co, Rh and Ir. Both Ce$_2$CoIn$_8$ and Ce$_2$RhIn$_8$ are
heavy-fermion superconductors. In the former indide, the ambient-pressure superconductivity
emerges at $T_{\rm c}$ = 0.4 K out of the paramagnetic normal state \cite{che_ce2coin8}, while in
Ce$_2$RhIn$_8$ an antiferromagnetic ordering below $T_{\rm N}$ = 2.8 K occurs at ambient
pressure, and the superconducting properties set in under applied pressure that suppresses the
magnetism and yields $T_{\rm c}$ = 2 K at $p$ = 2.3 GPa \cite{nic_ce2rhin8}. Ce$_2$IrIn$_8$, as the
only material from the series, does not exhibit any cooperative phase transition and remains a
heavy-fermion paramagnet down to 50 mK \cite{kim_ce2irin8}.

Very recently, we reported for the first time on the formation and the main physical properties of a
novel representative of the Ce$_2T$In$_8$ family with $T$ = Pd \cite{218_poly,218_sc}. Single
crystals of Ce$_2$PdIn$_8$ have been found to exhibit heavy fermion clean-limit
superconductivity below $T_{\rm c}$ = 0.68 K. In contrast, the compound studied in
polycrystalline form has not revealed any phase transition at low temperatures, and has been
characterized as a paramagnetic heavy-fermion system with a non-Fermi liquid character of its
electronic ground state.

In this paper, we communicate on the superconductivity in high quality polycrystals of
Ce$_2$PdIn$_8$. Based on the results of electrical resistivity and specific heat measurements
performed down to 350 mK, we provide evidence for the heavy-fermion superconducting state below
$T_{\rm c}$ = 0.7 K that emerges likely due to proximity of the system to an antiferromagnetic
quantum critical instability.

\section{Experimental details}

Polycrystalline sample of Ce$_2$PdIn$_8$ was synthesized by arc melting the stoichiometric
amounts of the
elemental components (Ce - 3N, Ames Laboratory, Pd - 3N, Chempur and In - 6N, Chempur) in a
copper-hearth furnace installed inside a glove-box filled with ultra-pure argon gas with
continuously controlled partial pressures of O$_2$ and H$_2$O to be lower than 1 ppm. The button was
flipped over and remelted several times to ensure good homogeneity. The weight losses after the
final melting were negligible (less than 0.2\%). Subsequently, the sample was wrapped with
tantalum foil, sealed an evacuated quartz tube and annealed at 700 $^o$C for 5 weeks.

Quality of the obtained alloy was checked by x-ray powder diffraction using an X'pert Pro
PanAnalytical diffractometer with CuK$_\alpha$ radiation and by energy dispersive x-ray (EDX)
analysis employing a Phillips 515 scanning electron microscope equipped with an EDAX PV 9800
spectrometer. Both techniques proved single-phase character of the sample, with the proper
stoichiometry and the primitive tetragonal crystal structure of the Ho$_2$CoGa$_8$-type (space
group \emph{P}4/\emph{mmm}). The structural refinement done using the program FULLPROF
\cite{fullprof} yielded the lattice parameters and the positional parameters very close to those
reported in Ref. \cite{218_poly}.

Magnetic measurements were carried out in the temperature range 1.71 - 400 K and in applied magnetic
fields up to 5 T using a Quantum Design superconducting quantum interference device (SQUID)
magnetometer. The heat capacity and the electrical resistivity were measured by the relaxation
method and the ac technique, respectively, over the temperature interval 350 mK - 300 K and in fields
up to 9 T employing a Quantum Design PPMS platform.

\section{Results and discussion}

All the main results of the present experimental study of polycrystalline Ce$_2$PdIn$_8$ are
summarized in the tree panels of Fig. 1. Apparently, the compound is a well localized magnetic
system with no long-range magnetic order down to the lowest temperatures studied. Above 70 K, the
magnetic susceptibility (see Fig. 1a) obeys the Curie-Weiss law with a large negative paramagnetic
Curie temperature $\theta_{\rm p} =  -41.8$ K and the effective magnetic moment $\mu_{\rm eff}$ =
2.41 $\mu_{\rm B}$ being close to that expected for trivalent Ce ions. At lower temperatures, an
upward deviation of the inverse magnetic susceptibility from the Curie-Weiss law is seen, which
likely manifests crystalline electric field (CEF) interactions. Below ca. 6 K, the magnetic
susceptibility shows an additional distinct upturn (see the inset to Fig. 1a), which may hint at
the presence of some critical spin fluctuations. The magnetization taken at 1.71 K is a linear
function of magnetic field without any hysteresis effect (not shown here), as expected for
paramagnetic state. In general, the new magnetic data of Ce$_2$PdIn$_8$ are nearly identical to
those reported before for the polycrystalline sample \cite{218_poly}.

As may be inferred from Fig. 1b, nearly the entire temperature dependence of the electrical
resistivity of Ce$_2$PdIn$_8$ is mainly governed by the Kondo effect (note the solid curve). Above
50 K, the $\rho(T)$ variation can be approximated by the formula
$\rho(T)=(\rho_0 + \rho_0^{\infty}) + c_{\rm ph}T + c_{\rm K} \ln{T}$, in which the first term
accounts for the conduction electrons scattering on static defects and disordered spins, the
second term represents the phonon contribution, whereas the third one results from the spin-flip
Kondo scattering. From the least-squares fitting one derives the parameters:
$\rho_0 + \rho_0^{\infty}$ = 127.5 $\mu \Omega$cm, $c_{\rm ph}$ = 0.11 $\mu \Omega$cm/K and
$c_{\rm K} = -18.6$ $\mu \Omega$cm. These values are almost identical to those derived
previously \cite{218_poly}.
The broad maximum in $\rho(T)$, seen around 30 K, can likely be associated with a transition from
incoherent to coherent Kondo regime. This crossover maximum occurs at slightly higher
temperature than that found for the polycrystal reported in Ref. \cite{218_poly}, which is in line
with higher metallurgical quality of the new sample (cf. the Summary). In the coherent state, the
resistivity rapidly decreases with decreasing temperature down to $T_{\rm c}$ = 0.7 K, at
which point an onset of the superconducting state is observed with a sharp resistivity drop to zero
value. Remarkably, above $T_{\rm c}$ the resistivity changes with temperature according to
the formula $\rho(T)=\rho_0 + aT^{\rm n}$ with
$\rho_0 = 6.4$ $\mu \Omega$cm, $a = 4.1$ $\mu \Omega$cm/K and $n$ = 1 (note the dashed line in Fig. 1b).
It is worthwhile emphasizing that this linear dependence of $\rho(T)$ is observed up to 9 K, i.e.
over more than a decade in the temperature. Such a distinct non-Fermi-liquid (NFL) character of the
resistivity is usually considered as a hallmark of quantum critical spin fluctuations in
two-dimensional systems with inherent antiferromagnetic correlations \cite{moriya}.

The superconducting transition at $T_{\rm c}$ = 0.7 K manifests itself as a sharp peak in the
low-temperature specific heat of Ce$_2$PdIn$_8$ (see Fig. 2c). At higher temperatures, $C(T)$
can be analyzed in terms of a sum of the electronic, Schottky and phonon contributions,
$C(T) = C_{\rm el} + C_{\rm Sch} + C_{\rm ph}$, where $C_{\rm el}$ has the NFL form
$C_{\rm el}(T) = aT\ln(T_0/T)$, $C_{\rm Sch}$ is written for a system of three Kramers
doublets and $C_{\rm ph}$ is represented by the Debye model. As shown in Fig. 2c, adopting the
values derived in Ref. \cite{218_poly} for the critical spin-fluctuation temperature ($T_0$ = 38
K) and for the energies of the excited crystal field levels ($\Delta_1$ = 60 K and $\Delta_2$ = 198 K),
and also taking the Debye temperature $\Theta_{\rm D}$ = 193 K, i.e. very close to that
estimated in Ref. \cite{218_poly} ($\Theta_{\rm D}$ = 184 K), one may reasonably well
describe the experimental $C/T$ data nearly in the entire temperature range above
$T_{\rm c}$.

The magnetic entropy, calculated from the excess specific heat due to the cerium 4$f$ electrons
(not shown here), reaches a value of $R\ln 2$ per Ce atom (expected for a doubly degenerated ground
state) not below 20 K. This distinct entropy reduction presumably results from Kondo screening
interactions. The Bethe Ansatz approach (for effective spin $s$ = 1/2) yields the Kondo
temperature of about 10 K, in perfect agreement with the value $T_{\rm K} \simeq
|\theta_{\rm p}/4| \approx$ 10 K, estimated from the magnetic susceptibility data shown. The
very same value of $T_{\rm K}$ was derived for Ce$_2$PdIn$_8$ in the previous studies
\cite{218_poly,218_sc}.

The low-temperature resistivity and the specific heat data of Ce$_2$PdIn$_8$ are shown in Figs. 2a
and 2b, respectively. Upon applying magnetic field the superconducting transition, defined as a
midpoint in $\rho(T)$ and as an inflection point above the peak in $C(T)$, shifts to lower
temperatures and gets suppressed below 0.35 K (the terminal temperature in this study) in a field of
2 T. In strong magnetic fields the $C/T$ ratio shows a tendency to saturate at a strongly enhanced
value $\gamma_{\rm n}$ of 1220 mJ/(mol K$^2$), thus proving heavy Fermi liquid character of
the compound studied. In zero field, the specific heat jump at $T_\textrm{c}$ = 0.7 K amounts to
about 1.5 J/(mol K$^2$), and hence the ratio $\Delta C/\gamma_{\rm n}T_{\rm c}$ is about
1.74, which is larger than a value of 1.43 predicted by the BCS theory.

The temperature dependence of the upper critical field $\mu_0 H_{\rm c2}$ is shown in Fig. 2c.
The initial slope $d\mu_0 H_{\rm c2}/dT$ attains as large value as -13.5 T/K, and
extrapolation to zero temperature yields $\mu_0 H_{\rm c2}(0)$ = 2.5 T. These parameters,
especially $\mu_0 H_{\rm c2}(0)$, are somewhat lower than those reported for
single-crystalline Ce$_2$PdIn$_8$, derived for the magnetic field applied along the $c$-axis
\cite{218_sc}. Nevertheless, their magnitudes clearly corroborate the heavy fermion character
of the superconducting state. In the framework of the model developed by Orlando \textit{et al.}
(Ref. \cite{orlando}), one may calculate the the effective mass $m^* \approx$ 193
$m_{\rm e}$, the electronic mean free path of the quasiparticles $l \approx$ 27 nm and the BCS
coherence length $\xi_0 \approx$ 4.9 nm. The so-derived relation $l \gg \xi_0$ clearly indicates
that Ce$_2$PdIn$_8$ exhibits clean limit superconductivity. Furthermore, from the formulas
given in Ref. \cite{gl}, one may also determine the Ginzburg-Landau coherence length
$\xi_{\rm LG}$ and the penetration depth $\lambda_{\rm GL}$ to be equal to 11.5 and 422 nm,
respectively. Hence, the Ginzburg-Landau parameter $\kappa_{\rm LG}$ is estimated to be
about 37, well within the range for type II superconductivity. In general, all these key
characteristics of the superconducting state in polycrystalline Ce$_2$PdIn$_8$ are similar to
those calculated for the single crystals (compare Table I in Ref. \cite{218_sc}). Some
differences in the two sets of the superconducting parameters should likely be attributed to
anisotropy in the electrical transport behavior, inherent to the tetragonal crystal structure of
the compound studied.

\section{Summary}
The experimental data obtained in this work undoubtedly demonstrate that high-quality
polycrystals of Ce$_2$PdIn$_8$ exhibit at low temperatures the heavy-fermion
superconductivity, which emerges out of the paramagnetic normal state with distinct non-Fermi
liquid character. This finding contrasts with the previous report on the properties of
polycrystalline Ce$_2$PdIn$_8$ that has been characterized as a paramagnet down to 0.35 mK. The
lack of superconductivity in the previously studied polycrystals can be rationalized by their
poorer metallurgical quality in comparison to the sample investigated in the present work.
Extreme sensibility of the superconducting properties to internal strains, structural disorder
and non-stoichiometry is a well known characteristic feature of unconventional coupling of
Cooper pairs (it is enough to recall here the case of CeCu$_2$Si$_2$ \cite{geg_cecu2si2}). The new
polycrystalline sample of Ce$_2$PdIn$_8$ was synthesized using higher-purity cerium (3N
instead of 99.8 wt.$\%$) in precisely controlled argon atmosphere (arc-furnace installed in an
glove-box) and then annealed at higher temperature (700 $^o$C viz. 600 $^o$C applied before) and
for longer time (five instead of four weeks). This procedure yielded the electrical resistivity of
ca. 10 $\mu \Omega$cm just above the onset of superconductivity in the new specimen, as compared to
the residual resistivity of about 40 $\mu \Omega$cm, reported in Ref. \cite{218_poly}.

The superconducting temperature measured for the present polycrystalline sample of
Ce$_2$PdIn$_8$ is equal to that reported in Ref. \cite{218_sc} for the single crystal. Also
the other main characteristics of the superconducting state are similar in both crystalline forms
of the compound. Thus, the new experimental data definitely corroborate all our previous
statements on the heavy-fermion superconductivity in Ce$_2$PdIn$_8$. On the other hand, the
present work does not support our arguments for the antiferromagnetic ordering above
$T_{\rm c}$. In contrast to the behavior of the single-crystalline sample studied in Ref.
\cite{218_sc}, the polycrystals of Ce$_2$PdIn$_8$ are paramagnetic in the entire normal
state. This apparent contradiction has recently been solved by Uhlirova \emph{et al.}
\cite{prague}, who demonstrated that the antiferromagnetism observed in as-grown single
crystals of Ce$_2$PdIn$_8$ is always due to small admixture of the impurity phase CeIn$_3$. Our own
on-going studies seem confirm the presence of a thin layer of the latter compound, sandwiched in
between single-crystalline slabs of the parent indide. Full account on the metallurgical
problems encountered in growing single crystals of Ce$_2$PdIn$_8$ will be given in our
forthcoming paper.

To conclude, Ce$_2$PdIn$_8$ is a novel heavy-fermion superconductor with $T_{\rm c}$ = 0.7 K
at ambient pressure. Its main superconducting parameters are fairly similar to those reported for
the closely related compounds CeCoIn$_5$, CeIrIn$_5$ and Ce$_2$CoIn$_8$
\cite{pet_cecoin5,pet_ceirin5,che_ce2coin8}. The non-Fermi liquid character of the normal
state hints at the presence of critical spin fluctuations of antiferromagnetic type. It seems very
likely that the superconductivity in Ce$_2$PdIn$_8$ emerges at the verge of underlying quantum
critical point instability, similarly to the case of the Ce$T$In$_5$ and Ce$_2T$In$_8$
relatives.

\section{Acknowledgement}
This work was supported by the Ministry of Science and Higher Education within the research
project No. N202 116 32/3270.





\bibliographystyle{elsarticle-num}
\bibliography{<your-bib-database>}



\end{document}